\documentclass[aip,apl,reprint]{revtex4-2}
\usepackage[utf8]{inputenc}
\usepackage{amsmath}
\usepackage{graphicx}
\usepackage{hyperref}

\begin{document}

\title{Nanocrystalline cubic boron arsenide growth on GaAs(100) by gas source molecular beam epitaxy}

\author{Matthew A. Delmont}
\affiliation{Department of Mechanical Engineering, University of California, Santa Barbara, California 93106, USA}

\author{Yu Wu}
\affiliation{Department of Electrical and Computer Engineering, University of California, Santa Barbara, California 93106, USA}

\author{Han-Hsuan Wu}
\affiliation{Department of Materials Science and Engineering, University of California, Irvine, California 92697, USA}

\author{Xiaoqing Pan}
\affiliation{Department of Materials Science and Engineering, University of California, Irvine, California 92697, USA}

\author{Christopher J. Palmstr{\o}m}
\affiliation{Department of Electrical and Computer Engineering, University of California, Santa Barbara, California 93106, USA}
\affiliation{Materials Department, University of California, Santa Barbara, California 93106, USA}

\author{Bolin Liao}
\email{bliao@ucsb.edu}
\affiliation{Department of Mechanical Engineering, University of California, Santa Barbara, California 93106, USA}

\date{\today}

\begin{abstract}
Cubic boron arsenide (c-BAs) is a promising ultrahigh thermal conductivity semiconductor, but its direct synthesis on lattice-mismatched substrates remains challenging. Here, we demonstrate the synthesis of nanocrystalline c-BAs thin films on GaAs(100) substrates using gas source molecular beam epitaxy. X-ray photoelectron spectroscopy reveals an optimal growth temperature window of \(250\text{--}400~^\circ\text{C}\) to maintain near stoichiometry and suppress severe arsenic loss. Surface and structural characterizations via atomic force microscopy, reflection high-energy electron diffraction, and grazing-incidence X-ray diffraction demonstrate a Volmer-Weber 3D island growth mode and a preferential crystallographic alignment with the GaAs substrate. High-resolution transmission electron microscopy confirms a chemically sharp interface without gallium interdiffusion into the c-BAs layer. These findings establish an initial step for the integration of c-BAs into advanced thermal and optoelectronic platforms.
\end{abstract}

\maketitle

The rapid miniaturization of electronic components has drastically increased power
densities, making high-efficiency thermal management one of the central bottlenecks in modern
semiconductor technology \cite{moore2014emerging}. On a macroscopic scale, the surge in artificial intelligence
infrastructure and data center operations requires intensive cooling systems that consume
massive local water and energy resources, creating pressing engineering and environmental
challenges \cite{siddik2021environmental}. Cubic boron arsenide (c-BAs) has emerged as a transformative candidate to address
these issues, boasting a high room-temperature thermal conductivity of \(1300-1500~\text{W/m}\text{K}\), or potentially even higher. \cite{tian2018unusual,kang2018experimental,lindsay2013first,niyikiza2025}. Beyond high thermal conductivity, c-BAs exhibits
critical operational advantages over diamond: its acoustic stiffness and lattice properties closely
match established semiconductors such as gallium arsenide (GaAs) and gallium nitride (GaN), drastically reducing thermal
expansion mismatches, interfacial phonon scattering, and contact thermal resistance \cite{kang2021}.

\begin{figure*}[htbp]
    \centering
    \includegraphics[width=0.75\textwidth]{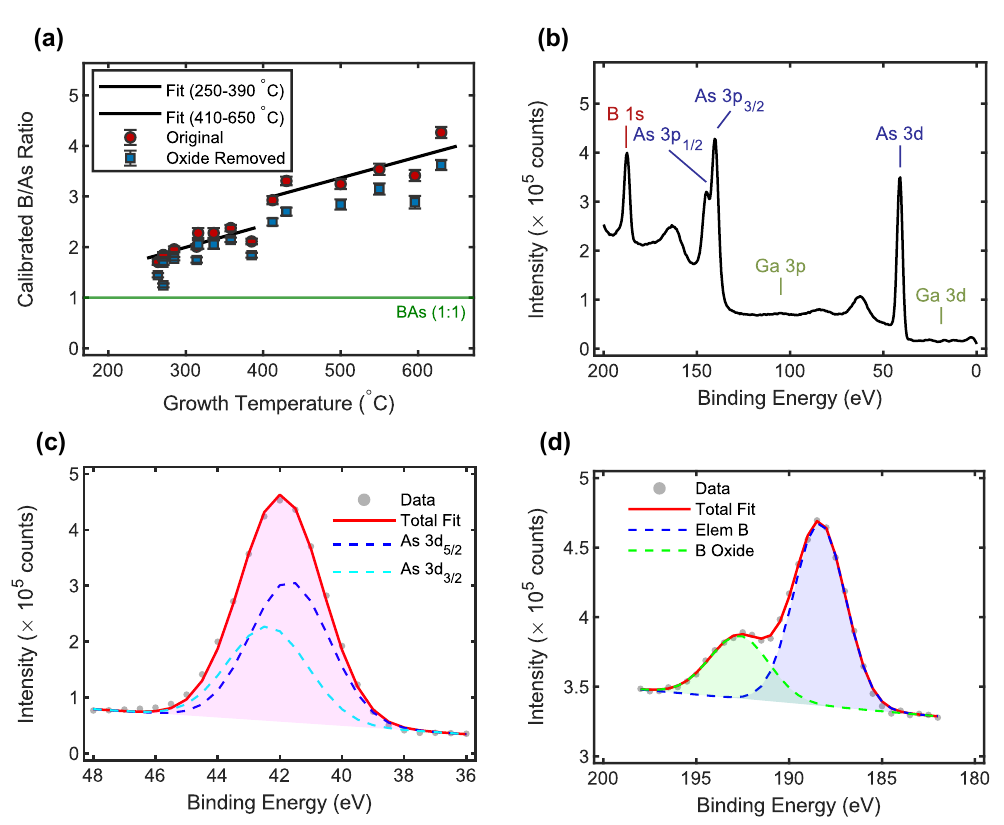}
    \caption{(a) Total boron to arsenic ratio of c-BAs films calibrated by a c-BAs single crystal compared to growth temperature measured by XPS. A relatively flat compositional plateau with excess boron is observed at \(250\text{--}400~^\circ\text{C}\), followed by a sharp loss of arsenic above \(400~^\circ\text{C}\). The ratio approaches the ideal c-BAs stoichiometry when surface boron oxide is removed by in-situ sputtering. (b) XPS survey spectrum of crystalline c-BAs film, demonstrating only boron and arsenic with no gallium detected from the substrate. (c) Fitting of As $3d$ XPS peak demonstrating no arsenic oxide. (d) Fitting of B $1s$ XPS peak revealing a boron oxide component due to preferential surface oxidation. The boron peak in boron arsenide is indistinguishable from elemental boron due to the highly covalent bonding.}
    \label{fig1}
\end{figure*}

Additionally, c-BAs possesses exceptional intrinsic electronic properties, including high
ambipolar carrier mobility and long photocarrier lifetimes, making it uniquely suited both as an
efficient thermal heat spreader and as an active standalone semiconductor for electronic and
photonic applications \cite{shin2022, choudhry2023}. Integrating c-BAs directly with GaAs provides an immediate
framework for high-performance optical sensors, detectors, and photovoltaics while
establishing a foundational platform for advanced heterostructure devices \cite{ren2026, bushick2020}.

Despite this potential, widespread deployment of c-BAs is severely bottlenecked by
growth difficulties driven by extreme disparities in melting temperatures, vapor pressures,
and atomic radii of boron and arsenic. Current bulk c-BAs crystals remain limited to small scales
(\(\sim 1~\text{mm}\)) with non-uniform structural quality, rendering commercial wafer-scale integration
unfeasible \cite{perri1958, lee2024, koirala2023}. Consequently, thin-film epitaxy represents the most practical pathway for
device integration; however, this approach is fundamentally constrained by a lack of lattice-matched substrates, including an over 15\% tensile lattice mismatch with GaAs(100). Such
extreme misfit strain typically destabilizes coherent film growth and induces high defect
densities \cite{matthews1974, evans1995}. Previous attempts at synthesizing c-BAs thin films have yielded mainly
amorphous structures with possibly thin epitaxial layers or boron subarsenide \cite{correia1986, cheng2025}. In this work, we demonstrate nanocrystalline boron arsenide films. Utilizing  low-temperature
seed initiation followed by temperature ramping to promote 3D island coalescence, we
provide a route for integrating c-BAs into advanced thermal and optoelectronic
platforms.
\vspace{1em}

Cubic boron arsenide (c-BAs) thin films were grown on GaAs(100) substrates in a
chemical beam epitaxy system with an ultra-high vacuum environment (base pressure \(< 10^{-10}~\text{Torr}\)). High-purity elemental boron and thermally cracked arsine (\(\text{AsH}_3\) providing
an \(\text{As}_2\) molecular flux) were used as sources of boron and arsenic respectively.
\begin{figure*}[htbp]
    \centering
    \includegraphics[width=0.75\textwidth]{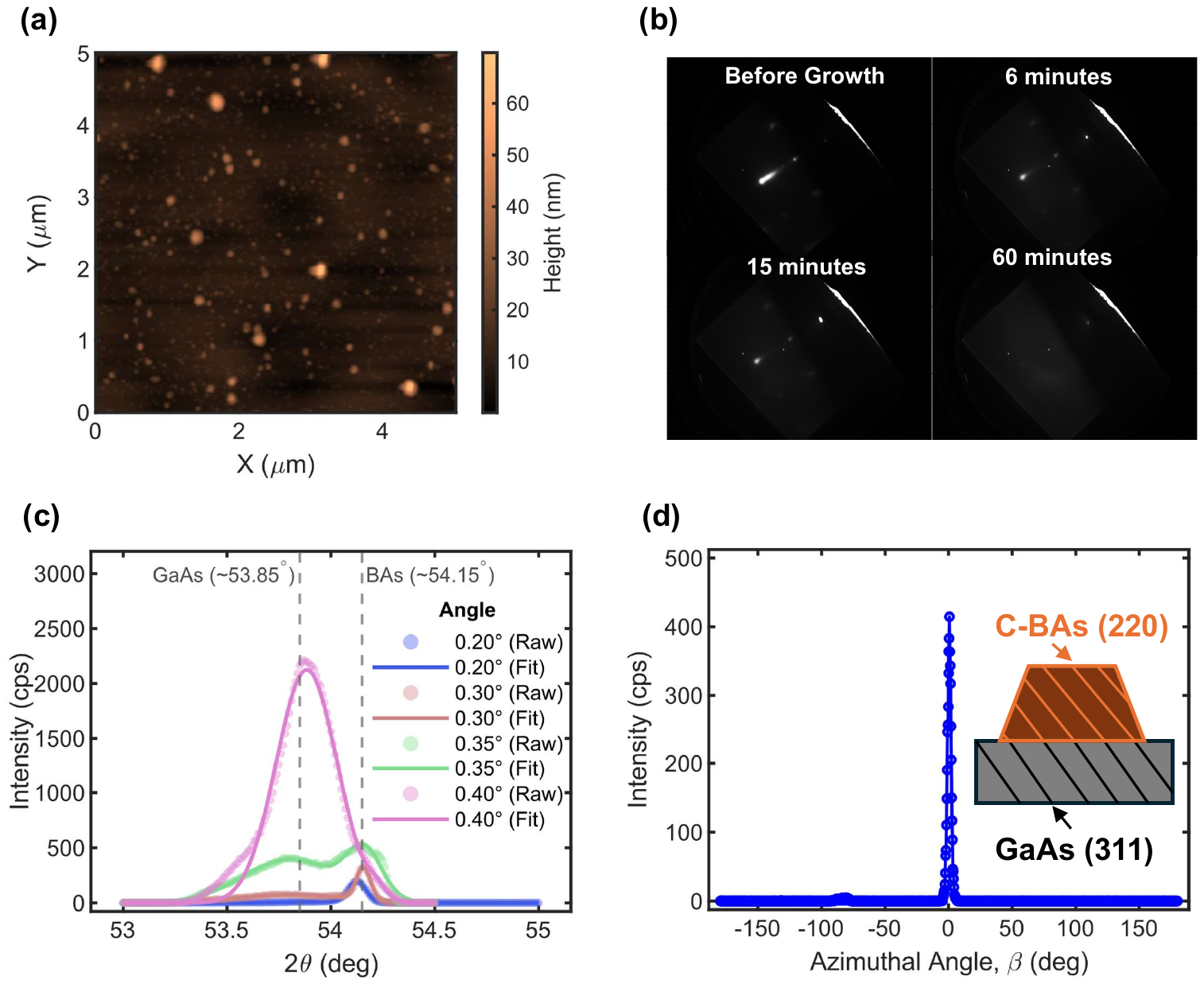}
    \caption{(a) Atomic force microscopy (AFM) image of BAs surface displaying surface islands. (b) Reflection high-energy electron diffraction (RHEED) patterns during different times of c-BAs growth demonstrating surface roughening and island growth mode. (c) Grazing-incidence X-ray diffraction scan at different incident angles (\(0.2^\circ\) to \(0.4^\circ\)). Below the critical angle for GaAs (\(\sim 0.3^\circ\)), only the c-BAs(220) peak reflection is visible. Above the critical angle, the GaAs substrate (311) peak dominates. (d) Azimuthal scan of the BAs(220) reflection showing preferential orientation and alignment with the GaAs substrate, along with inset diagram picturing this}
    \label{fig2}
\end{figure*}
Following native oxide desorption and the deposition of a GaAs buffer layer, c-BAs growth was carried out under As-rich flux conditions, with a 130:1 arsenic to boron atomic flux ratio, at effective substrate temperatures initially at \(300~^\circ\text{C}\) ramping up to \(360~^\circ\text{C}\) for the films analyzed in detail. The nominal film growth rate was approximately \(\sim 9~\text{nm/h}\). Films at other growth temperatures were also analyzed by x-ray photoelectron spectroscopy (XPS) to measure the stoichiometry and determine the c-BAs growth window.

XPS measurements were taken with a ThermoFisher Escalab Xi+ instrument with an aluminum k$_\alpha$ source emitting x-rays with a characteristic energy of 1486.68 eV. Argon sputtering was performed to remove native carbon and oxide layers formed upon exposure to air with an energy 2000 eV and a time of up to 10 seconds with a calibrated removal of tantalum oxide of 0.4 nm/sec.

Surface morphology was evaluated using an Asylum MFP-3D atomic force microscope (AFM) in AC air tapping mode with Bruker TESPA-V2 AFM tips. Reflection high-energy electron diffraction (RHEED) measurements were performed in-situ with an accelerating voltage of 11 kV and an emission current of 0.2 mA.

Grazing-incidence x-ray diffraction (GIXRD) measurements were conducted on a
Rigaku SmartLab High-Resolution Diffractometer equipped with a HyPix-3000 detector
using a Cu K\(\alpha\) source (\(\lambda = 1.5406~\text{\AA}\)) \cite{marra1979}. Samples were mounted onto fused quartz substrates
via indium bonding. Incident beam divergence was restricted using a \(0.1~\text{mm}\) computer-controlled primary slit and a \(5~\text{mm}\) manual slit to prevent x-ray spillover on the \(3 \times 2~\text{cm}^2\) sample at shallow grazing angles. Detector acceptance slits were set to \(1~\text{mm}\) to resolve
the closely spaced diffraction peaks. X-ray reflectivity (XRR) determined the film's critical angle
for total external reflection (\(\alpha_c\)) to be \(\sim 0.30^\circ\) \cite{sinha1988} [Fig. S1(a)].

Cross sectional transmission electron microscopy (TEM) samples were prepared utilizing focused ion beam milling. The measurements were performed on JEOL Grand ARM ADF-TEM/STEM with accelerating voltages of 300 kV, along with electron energy loss spectroscopy (EELS). Additional TEM samples were measured on the Thermo Scientific Spectra 200 with accelerating voltages of 200 kV, along with Energy-Dispersive X-ray Spectroscopy (EDS) measurements.

\begin{figure*}[htbp]
    \centering
    \includegraphics[width=0.75\textwidth]{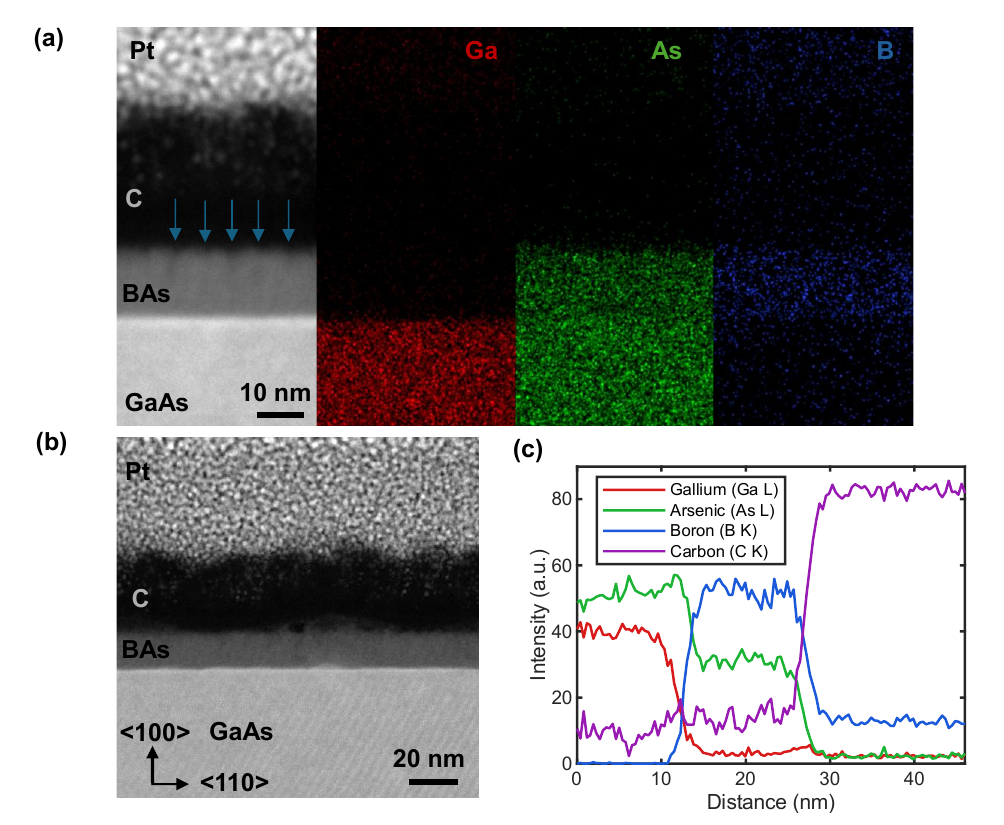}
    \caption{(a) High-angle annular dark-field transmission electron microscopy (HAADF-TEM) image with arrows marking possible grain boundary regions. Energy-dispersive X-ray spectroscopy (EDS) mapping revealing elemental transition between GaAs substrate and BAs film. (b) High-resolution HAADF-TEM image of the BAs-GaAs interface. (c) Electron energy loss spectroscopy (EELS) profile across the BAs-GaAs interface demonstrating a sharp chemical transition and uniform BAs film composition.}
    \label{fig3}
\end{figure*}
\vspace{1em}
Previous studies on the thin-film growth of cubic boron arsenide (c-BAs) lacked long-range structural order and observed a loss of arsenic at elevated growth temperatures \cite{cheng2025}.
Single-crystal synthesis has similarly demonstrated a thermodynamic preference for forming the
boron subarsenide phase at higher temperatures \cite{ziyaee2019}. 
XPS measurements delineate the growth temperature window for the c-BAs phase. In the temperature range of \(250~^\circ\text{C}\) to \(400~^\circ\text{C}\), a small slope in arsenic concentration
with respect to temperature is observed. At close to \(250~^\circ\text{C}\), the film approaches 1:1 stoichiometry. Above \(\sim 400~^\circ\text{C}\), a sharp transition occurs,
characterized by a rapid loss of arsenic as the film composition trends toward boron subarsenide, similar as seen in previous reports \cite{cheng2025}
[Fig.~\ref{fig1}(a)]. Furthermore, films deposited in this higher temperature regime generally lack
long-range structural order and have extremely rough pitting surface features [Fig. S2].

Compared to stoichiometric c-BAs single crystals, the grown c-BAs films exhibit a lower
overall arsenic content. This deviation may be partially attributed to XPS measurement artifacts, such as surface topology
interactions and matrix effects, which also underestimated the arsenic concentration in high-quality single-crystal reference samples, yielding a measured B/As ratio of 1.1. [Fig S3.(a)] \cite{shard2020}.
Additionally, peak fitting reveals that ambient air exposure leads to the formation of a surface
boron oxide layer [Fig.~\ref{fig1}(d)]. Because boron has a strong thermodynamic potential for oxidation
while arsenic oxides are volatile at elevated temperatures, preferential boron oxidation and potential surface segregation
can occur, artificially elevating the measured surface boron content relative to the film bulk \cite{haynes2011}.
When this boron oxide is removed by argon sputtering and peak fitting, the relative ratio
decreases to \(\sim 1.2\). This excess boron can potentially be attributed to the preferential sputtering of arsenic
compared to boron, which increased the relative ratio in boron arsenide single crystals \cite{singer1981, kelly1985} [Fig. S3(b)]. Alternatively, regions of amorphous boron within the film
matrix may also contribute to the observed excess boron, as commonly seen in other boron compound films \cite{domnich2011,ahn2025}.

Crucially, XPS survey spectra reveal no detectable gallium peaks, confirming that the c-BAs film forms a continuous and coherent layer over the GaAs substrate [Fig.~\ref{fig1}(b)]. The measured arsenic must also originate only from the c-BAs film because the film thickness is greater than the photoelectron attenuation length. This is further demonstrated by the lack of gallium signal, with gallium 3d photoelectrons having a greater penetration depth than arsenic 3d photoelectrons\cite{seah1979}. The following sections analyze in detail a c-BAs film grown initially at \(300~^\circ\text{C}\) and heated up to \(360~^\circ\text{C}\) during growth.
\vspace{1em}

  Large-area topographic scans [Fig.~\ref{fig2}(a)] reveal circular islands of heights up to 70~nanometers with a root-mean-square roughness of 5~nanometers. Additional AFM scans and scanning electron microscopy images demonstrate similar circular island structures in different locations on the film, demonstrating that these features are consistent across the film [Fig. S4]. The in-situ reflection high-energy electron diffraction (RHEED) performed during growth similarly demonstrates a roughness transition. The non-principal azimuth RHEED signal shows a central streak initially associated with the GaAs substrate that quickly transforms into an array of dots, demonstrating a roughening that can be attributed to a 3D growth mode [Fig.~\ref{fig2}(b)]. This morphological evolution characterized by initial three-dimensional (3D) island nucleation followed by lateral coalescence is evidence of a
Volmer-Weber growth mode \cite{volmer1926}. This growth pathway is energetically favored due to the large
lattice mismatch (\(\sim 15.5\%\)) between the c-BAs film (\(a = 4.777~\text{\AA}\)) and the GaAs substrate (\(a = 5.653~\text{\AA}\)). The substantial interfacial strain energy suppresses two-dimensional layer-by-layer
(Frank-van der Merwe) growth in favor of 3D island formation \cite{volmer1926, frank1949}.

\vspace{1em}
\vspace{1em}
The structural relationship between the c-BAs film and the GaAs(100) substrate was
investigated using coplanar vertical \(2\theta\) scans [Fig.~\ref{fig2}(c)] at incident angles (\(\alpha_i\)) above and
below \(\alpha_c\). For \(\alpha_i < \alpha_c\) (e.g. at \(0.20^\circ\)) the X-ray beam was confined to the surface region, revealing the
surface-sensitive c-BAs(220) reflection at \(2\theta \approx 54.27^\circ\) \cite{lee2024}. Increasing the incident angle
above \(\alpha_c\) (\(\alpha_i \ge 0.35^\circ\)) enabled penetration into the substrate, introducing the
adjacent GaAs(311) reflection at \(2\theta \approx 53.73^\circ\) \cite{blakemore1982}. In this coplanar geometry, setting \(2\theta \approx 53.7^\circ\) at a shallow incident angle (\(\alpha_i \approx 0.30^\circ\)) tilts the net scattering vector \(\vec{q}\) by an angle \(\gamma = \theta - \alpha_i \approx 26.56^\circ\) relative to the [100] surface normal. This orientation aligns closely with the
theoretical inclination of the GaAs(311) planes (\(25.24^\circ\)), enabling the substrate reflection to be
detected once the beam penetrates beyond \(\alpha_c\). Conversely, for an ideal \((100)\)-oriented cubic crystal (\(\text{c-BAs}(100) \parallel \text{GaAs}(100)\)), the theoretical interplanar angle between the \([100]\) surface normal and the \((220)\) plane normal is \(\phi_{\text{BAs}(220)} = 45.00^\circ\). Detecting the c-BAs(220) reflection at a scattering vector inclination of \(\gamma \approx 26.84^\circ\) indicates an effective out-of-plane crystallite tilt of:
\begin{equation}
\Delta\phi = \phi_{\text{BAs}(220)} - \gamma \approx 45.00^\circ - 26.84^\circ = 18.16^\circ \approx 18.2^\circ.
\end{equation}
This \(\sim 18.2^\circ\) tilt reveals that strain relaxation during Volmer-Weber 3D island growth induces significant local crystallite misorientation, domain tilting, and faceted island growth under the \(\sim 15.5\%\) compressive lattice mismatch. 

To confirm in-plane orientation, an azimuthal rotation scan (\(\beta\)) was recorded with the $2\theta$ fixed at 54.15 degrees [Fig.~\ref{fig2}(d)]. A single narrow diffraction peak centered at \(\beta = 0^\circ\) is observed, with an absence of background intensity across the remaining rotational
span. This azimuthal confinement, with a full-width half maximum (FWHM) of  $3.4^\circ$, confirms that the c-BAs film is highly oriented in-plane
rather than randomly polycrystalline. The concurrent alignment of both
the c-BAs(220) film and GaAs(311) substrate projections along this identical azimuthal
orientation suggests preferential texturing of the film with respect to the substrate. To evaluate the out-of-plane mosaic spread of the BAs islands while maintaining grazing incidence conditions, an out-of-plane rocking curve ($\chi$-scan) was performed across the $54.15^\circ$ reflection. The out-of-plane misorientation distribution yields a FWHM of $3.5^\circ$, indicating significant mosaic tilt among the islands [Fig. S1(b)].

Taken together, these results indicate that, under the large overall lattice
mismatch ($15.5\%$, with c-BAs the smaller lattice), Volmer--Weber
three-dimensional island growth proceeds by nucleation of crystallites in a
single tilted, low-misfit orientation [Eq.~(1)], rather than by
cube-on-cube epitaxy. The finite out-of-plane mosaicity reflects local
misorientation and faceting of the coalescing islands.

\vspace{1em}
To directly evaluate the cross-sectional morphology, interfacial abruptness, and spatial uniformity of the c-BAs/GaAs heterostructure, high-angle annular dark-field transmission electron microscopy (HAADF-TEM) was conducted on the coalesced regions between the isolated islands. The cross-sectional HAADF-TEM image along with the EDS elemental maps confirms the presence of a continuous, uniform BAs film with a thickness of \(\sim 15\text{--}20~\text{nm}\) beneath the protective carbon and platinum capping layers. Furthermore, the columnar structural appearance of the film at the interface between the protective carbon cap and the boron arsenide suggests a nanocrystalline microstructure [Fig.~\ref{fig3}(a)] \cite{petrov2003}. The EDS maps display clear spatial partitioning of gallium in the substrate and boron in the film layer, while arsenic distribution spans across both sides of the interface. 

High-resolution HAADF-TEM imaging of the heterointerface [Fig.~\ref{fig3}(b)] demonstrates a well-defined boundary between the c-BAs layer and the underlying GaAs(100) substrate. To quantitatively examine chemical abruptness and check for interfacial interdiffusion, an EELS elemental profile was recorded across the heterostructure [Fig.~\ref{fig3}(c)]. The EELS intensity profiles for Ga \(L\), As \(L\), B \(K\), and C \(K\) edges reveal a sharp transition at the BAs/GaAs boundary (\(\sim 12~\text{nm}\) position). The Ga \(L\) signal drops sharply to baseline at the interface, while the B \(K\) signal rises rapidly and remains constant across the \(\sim 16~\text{nm}\) film thickness before terminating at the protective carbon capping layer (\(\sim 28~\text{nm}\)).
The complete absence of gallium within the BAs film independently confirms the XPS survey findings (Fig.~\ref{fig1}), verifying that the c-BAs growth forms a continuous, unmixed film over the GaAs substrate despite the significant lattice mismatch. The lattice structure at the interface is evaluated in [Fig. S5], but the structures at the interface cannot be conclusively determined as c-BAs or GaAs.

While direct molecular beam epitaxy of c-BAs on highly mismatched GaAs substrates represents a key initial step toward scalable thin-film integration, bridging the gap from initial nucleation to practical device implementation requires further growth optimization. Overcoming grain boundary scattering and defect creation will be vital to realizing the high thermal conductivity inherent to c-BAs in thin-film architectures \cite{cahill2003nanoscale}. With continued advances in interfacial engineering, stoichiometry control, and microstructural tailoring, c-BAs heterostructures hold significant promise as integrated heat spreaders and active semiconductor layers for next-generation thermal management and optoelectronic applications.

\section*{Author Declarations}
A United States provisional patent about the results in this paper has been filed.

\section*{Data Availability}
The data that supports the findings of this study are available from the corresponding author upon reasonable request.

\section*{Acknowledgments}
This work is supported by the U.S. National Science Foundation Future of
Semiconductors (FuSe) Program under the award number DMR-2425439. The research reported here made use of the Microscopy and Microanalysis Facility (MMF), as well as the X-ray Facility, at the University of California, Santa Barbara (ROR:03f1rr013) and the MRL Shared Experimental Facilities that are supported by the MRSEC Program of the NSF under Award No.DMR 2308708; a member of the NSF-funded Materials Research Facilities Network. The authors acknowledge the use of facilities and instrumentation at the UC Irvine Materials Research Institute (IMRI) supported in part by the National Science Foundation Materials Research Science and Engineering Center program through the UC Irvine Center for Complex and Active Materials (DMR-2011967).
\bibliography{references}

\end{document}


\title{Supplementary Information: Nanocrystalline cubic boron arsenide growth on GaAs(100) by gas source molecular beam epitaxy}
\author{Matthew A. Delmont}
\affiliation{Department of Mechanical Engineering, University of California, Santa Barbara, California 93106, USA}

\author{Yu Wu}
\affiliation{Department of Electrical and Computer Engineering, University of California, Santa Barbara, California 93106, USA}

\author{Han-Hsuan Wu}
\affiliation{Department of Materials Science and Engineering, University of California, Irvine, California 92697, USA}

\author{Xiaoqing Pan}
\affiliation{Department of Materials Science and Engineering, University of California, Irvine, California 92697, USA}

\author{Christopher J. Palmstr{\o}m}
\affiliation{Department of Electrical and Computer Engineering, University of California, Santa Barbara, California 93106, USA}
\affiliation{Materials Department, University of California, Santa Barbara, California 93106, USA}

\author{Bolin Liao}
\email{bliao@ucsb.edu}
\affiliation{Department of Mechanical Engineering, University of California, Santa Barbara, California 93106, USA}

\date{\today}
\maketitle

\begin{figure*}[p]
    \centering
    \includegraphics[width=0.85\textwidth, height=0.75\textheight, keepaspectratio]{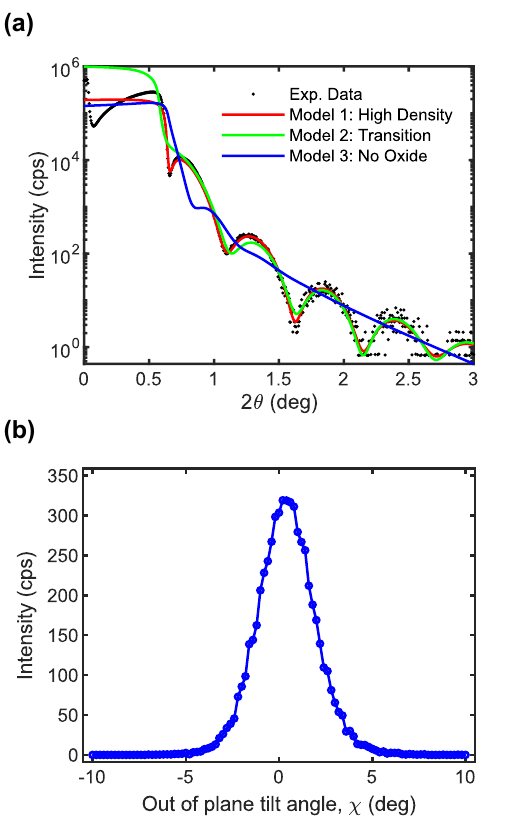}
    \caption{a) X-ray reflectivity measurement of c-BAs film grown at \(300~^\circ\text{C}\)-\(360~^\circ\text{C}\) for two hours. The fitting demonstrates that the film consists of a approximately 16 nm c-BAs layer and a 2nm boron oxide top layer. A model with a transition region between the oxide and the c-BAs, as well as a model with no oxide failed to match the XRR measured. Therefore, the film consists of a c-BAs layer with a sharp interface with a boron oxide layer. b) Rocking curve measurement of c-BAs film in the $\chi$ direction, while maintaining a grazing incidence of 0.2 degrees. The FWHM of the peak is 3.5 degrees demonstrating wide mosaic spread of grain orientation angles in the film.}
    \label{SI_fig1}
\end{figure*}

\begin{figure*}[p]
    \centering
    \includegraphics[width=0.85\textwidth, height=0.75\textheight, keepaspectratio]{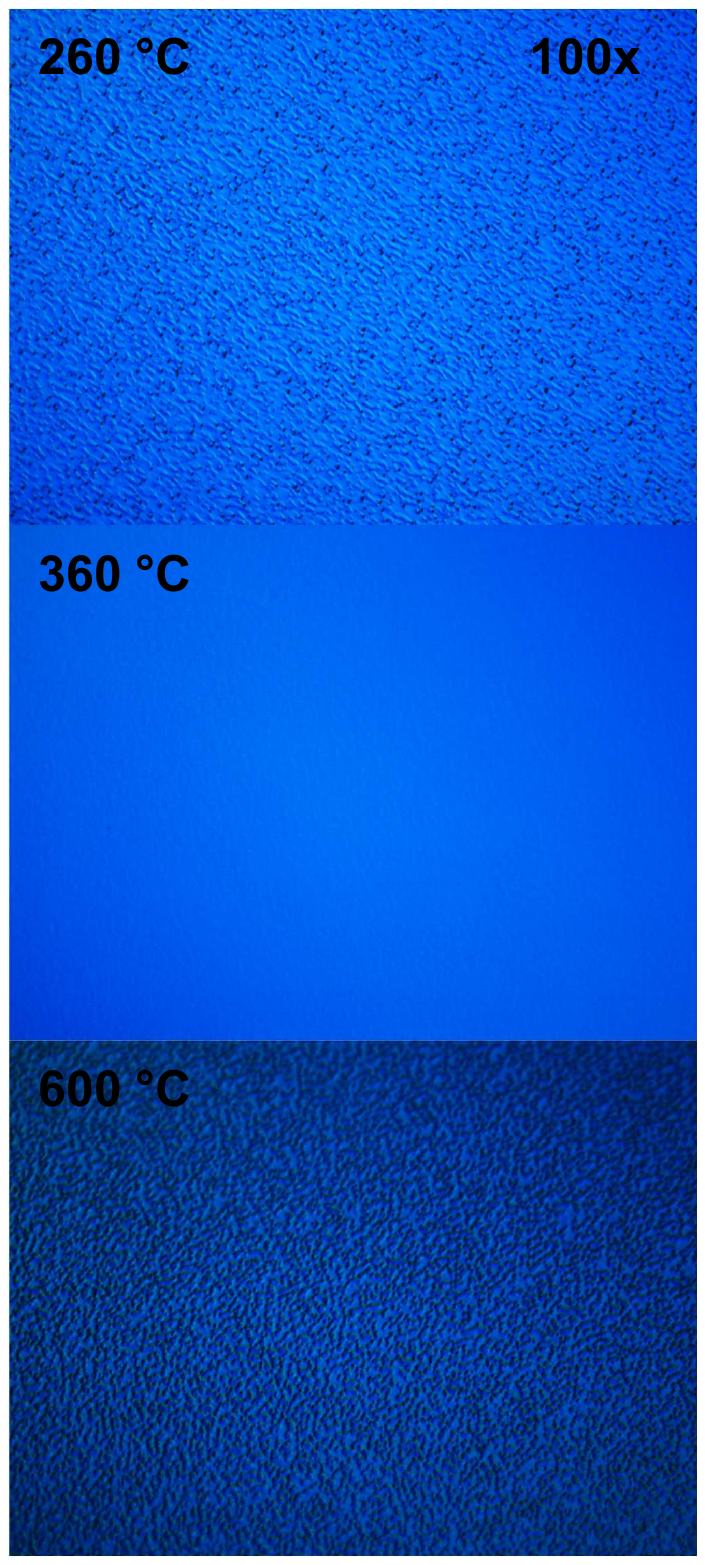}
    \caption{Optical microscope images at 100x magnification of the surface of samples grown at different temperatures. At \(260~^\circ\text{C}\), the surface is characterized by a wormy texture characteristic of low adatom mobility. At \(360~^\circ\text{C}\), the film is smooth optically. At \(600~^\circ\text{C}\), the sample surface is covered in round pits, likely due to arsenic desorption at this temperature.}
    \label{SI_fig2}
\end{figure*}

\begin{figure*}[p]
    \centering
    \includegraphics[width=0.85\textwidth, height=0.75\textheight, keepaspectratio]{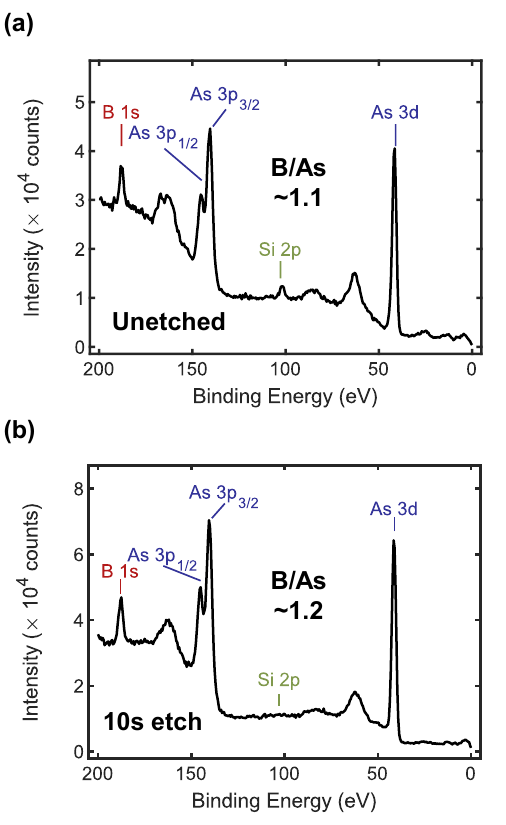}
    \caption{ a) XPS measurement of boron arsenide single crystal, demonstrating a surface arsenic deficiency with a boron to arsenic ratio of approximately 1.1. A silicon oxide peak is noticeable near 104 eV binding energy b) XPS measurement of boron arsenide single crystal after sputtering for 10 seconds with a loss of arsenic increasing the boron to arsenic ratio to 1.2, as well as removing the silicon oxide layer.}
    \label{SI_fig3}
\end{figure*}


\begin{figure*}[p]
    \centering
    \includegraphics[width=0.85\textwidth, height=0.75\textheight, keepaspectratio]{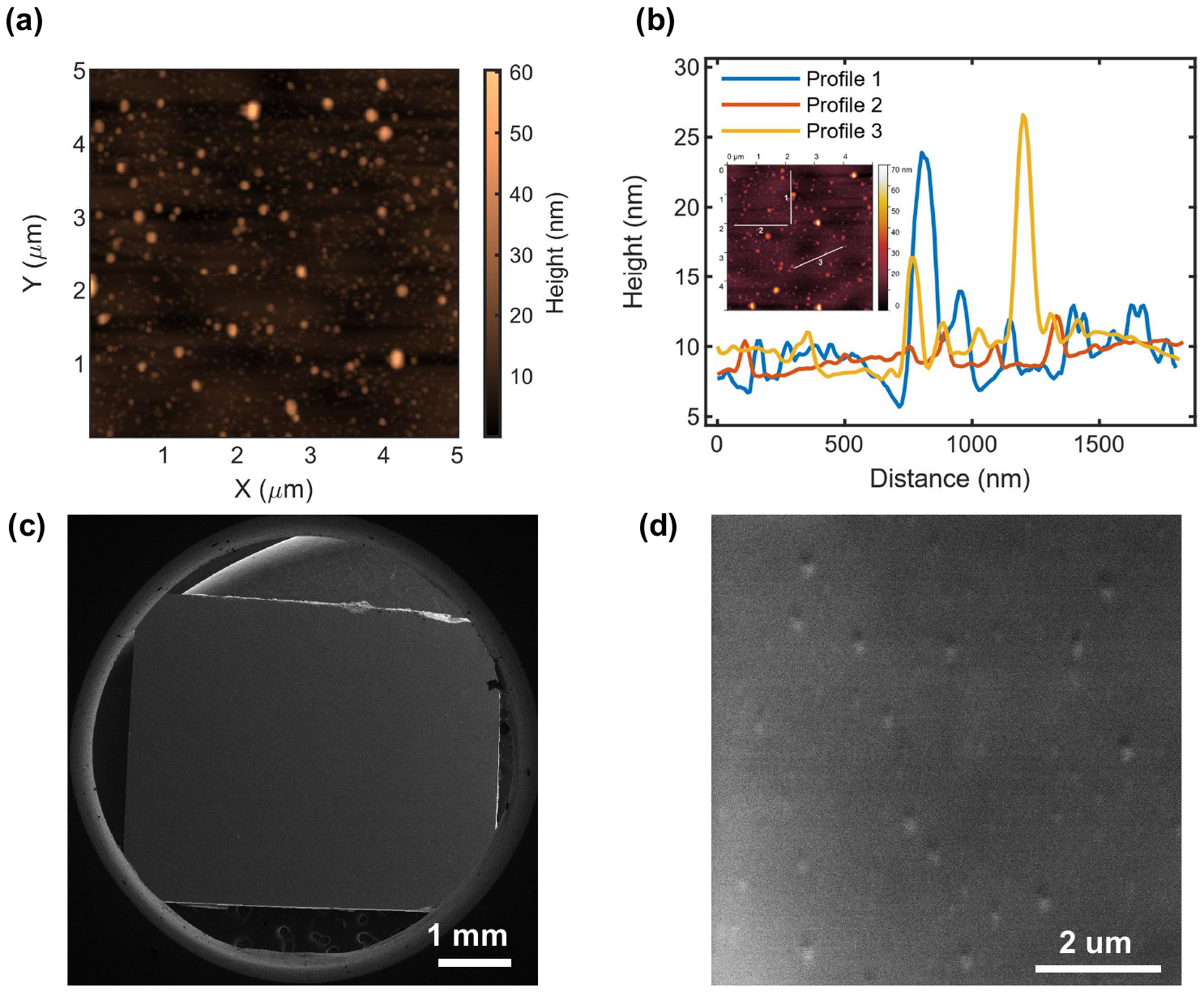}
    \caption{a) 5x5 um AFM image of second region of the sample grown at \(300~^\circ\text{C}\) up to \(360~^\circ\text{C}\) b) Line cuts of the same AFM scan demonstrating the variety of island sizes on the surface, as well as the broad slopes of the island sidewalls. c) Low magnification SEM image demonstrating macroscopic smoothness of the film d) High magnification SEM image showing the islands of 100s of nanometers in width, similar to the AFM images. }
    \label{SI_fig4}
\end{figure*}
\begin{figure*}[p]
    \centering
    \includegraphics[width=0.85\textwidth, height=0.75\textheight, keepaspectratio]{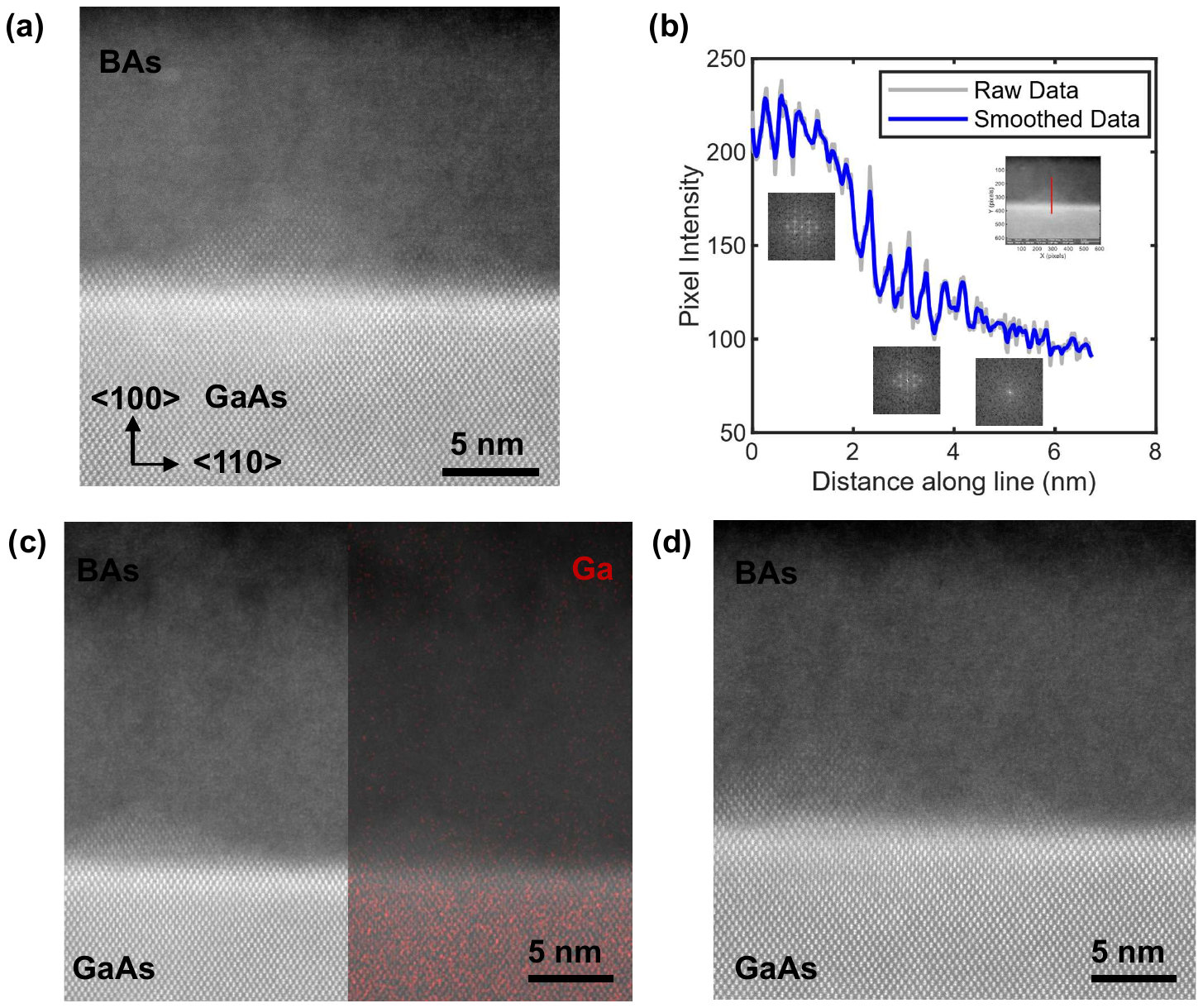}
    \caption{a) High resolution HAADF-TEM image of pyramidal structure at the BAs-GaAs interface b) Structural analysis of the transition zone reveals that regular atomic spacing remains even when intensity has decreased. Intensity in HAADF increases with atomic mass, suggesting a lower mass crystalline region at the interface. However, the difference in lattice between these structures and the GaAs is indistinguishable from FFT c) EDS measurements reveal a lack of gallium in the structures seen at the interface d) Additional structures at the BAs-GaAs interface from other locations. }
    \label{SI_fig5}
\end{figure*}